\pdfoutput=1
\documentclass[11pt,fleqn]{article}
\usepackage[T1]{fontenc}
\usepackage{lmodern}
\usepackage{placeins}
\usepackage{tikz}
\usetikzlibrary{arrows.meta,positioning,fit,backgrounds}
\usepackage{amsmath,amssymb,amsthm,mathtools}
\usepackage[margin=1in]{geometry}
\usepackage{xcolor}
\usepackage{enumitem}
\usepackage{algorithm,algpseudocode}
\usepackage[hypertexnames=false,bookmarksnumbered=true,final]{hyperref}
\usepackage[capitalize,sort]{cleveref}

\definecolor{textRed}{RGB}{145,35,45}
\definecolor{textBlue}{RGB}{35,75,180}
\definecolor{datablue}{RGB}{35,75,180}
\definecolor{datateal}{RGB}{20,112,92}
\definecolor{datared}{RGB}{200,74,88}
\definecolor{datagray}{RGB}{85,88,96}
\hypersetup{colorlinks,linkcolor=textRed,citecolor=textRed,urlcolor=textBlue,
  pdftitle={The Complexity of Fair and Stable Data Exchange},
  pdfauthor={Hannaneh Akrami, Bhaskar Ray Chaudhury}}

\newcommand{\eps}{\varepsilon}
\newcommand{\PPAD}{\mathsf{PPAD}}

\newcommand{\FP}{\mathsf{FP}}
\newcommand{\Lip}{\operatorname{Lip}}
\newcommand{\clp}{\operatorname{clip}}
\newcommand{\norminf}[1]{\left\lVert#1\right\rVert_\infty}
\newcommand{\one}{\mathbf 1}
\tikzset{>=Latex,
 agent/.style={rectangle,draw=datared,fill=datared!3,line width=.7pt,minimum size=6mm,inner sep=2pt},
 fixedagent/.style={agent,draw=datablue,fill=datablue!7},
 dataedge/.style={->,draw=black,line width=.75pt},
 targetedge/.style={->,draw=datateal,line width=1pt},
 slackedge/.style={->,draw=datablue,dashed,line width=.9pt},
 elabel/.style={fill=white,inner sep=1.3pt,font=\small},
 every picture/.style={line cap=round,line join=round}}

\theoremstyle{plain}
\newtheorem{theorem}{Theorem}[section]
\newtheorem{lemma}[theorem]{Lemma}
\newtheorem{proposition}[theorem]{Proposition}
\newtheorem{corollary}[theorem]{Corollary}
\theoremstyle{definition}
\newtheorem{definition}[theorem]{Definition}

\title{The Complexity of Fair and Stable Data Exchange}
\author{Hannaneh Akrami\textsuperscript{1,2} and Bhaskar Ray Chaudhury\textsuperscript{3}\\[1ex]
{\small \textsuperscript{1}Max Planck Institute for Informatics, Saarland Informatics Campus, Saarbr\"ucken, Germany}\\
{\small \textsuperscript{2}Hertz Chair for Algorithms and Optimization, University of Bonn}\\
{\small \textsuperscript{3}University of Illinois Urbana-Champaign}}
\date{}

\begin{document}
\pagenumbering{Alph}
\maketitle
\thispagestyle{empty}

\begin{abstract}
Data exchange is a fundamental challenge for Web applications such as recommender systems, personalized search, and fraud detection, where combining data across participants can improve predictions by expanding coverage and revealing patterns that individual datasets miss. Sustaining such collaboration requires fair returns for participants' contributions and protection against groups leaving to pursue a more beneficial exchange.
These requirements are formally captured by \emph{reciprocity}, which equates each participant's received utility with its credited contribution to others, and \emph{core stability}, which rules out any coalition exchanging its own data to make every member strictly better off. Akrami et al.~\cite{Akrami2025} prove that such exchanges exist under broad assumptions, but leave the study of efficient algorithms to determine such an exchange unresolved. Two fundamental cases are submodular utilities, which capture diminishing returns from datasets with overlapping information, and supermodular utilities, which capture complementarities between datasets.
We settle the approximate computation of reciprocal and core-stable exchanges in both cases, establishing a computational separation under Shapley sharing. For cross-monotone shares, which include Shapley shares under monotone submodular utilities, we give a circulation algorithm that computes an exactly core-stable, $\eps$-reciprocal exchange with efficient access to approximate shares. Its running time is $\textup{poly}(n,1/\eps,L)$, where $n$ is the number of participants, and $L$ is a Lipschitz bound on how each donor's credit changes with its own contribution. This yields an FPTAS when $L$ is polynomially bounded. In contrast, finding an $\eps$-reciprocal, $\eps$-core-stable exchange is PPAD-hard for monotone lattice-supermodular utilities, even at inverse-polynomial accuracy, with constant $L$ and exact polynomial-time evaluation of Shapley shares. Therefore, we settle the open problems in Akrami et al.\ by showing that overlapping information permits efficient fair and stable exchange, while complementarities create a computational barrier even when contribution credits are easy to evaluate.

\vspace{0.3cm}

\noindent\emph{The algorithm and hardness result were developed through active interactions with ChatGPT 6.0 (Astra); the authors verified and simplified the proofs and prepared the final exposition.}
\end{abstract}

\clearpage
\pagenumbering{arabic}
\tableofcontents
\clearpage

\section{Introduction}
\label{sec:introduction}

\paragraph{Data exchange.}
Data fuels modern AI and machine learning: access to large, informative datasets has enabled advances in statistical learning and Web applications~\cite{Halevy2009}. Much of this data is distributed across organizations, making exchanges of useful data valuable to participants with different learning tasks. Yet a data-exchange platform cannot rely on the benefits of collaboration alone. Data holders expect a return commensurate with their contributions~\cite{OECD2019}, and lack of reciprocity has been documented as a barrier to sharing public-health data~\cite{VanPanhuis2014}. We study data exchange without monetary transfers, where each agent's utility depends on the data it receives from others. Following Bhaskara et al.~\cite{Bhaskara2024} and Akrami et al.~\cite{Akrami2025}, we require an exchange to satisfy \emph{reciprocity}: each agent's received utility equals the credit it earns through its contributions to others. Reciprocity alone permits the trivial outcome in which no data is shared. We therefore also require \emph{core stability}: no coalition can strictly improve every member's utility by exchanging its own data internally. Such a deviating exchange need not itself satisfy reciprocity.

\paragraph{Existence of fair and stable exchange.}
Akrami et al.~\cite{Akrami2025} establish existence for all continuous, monotone utilities and continuous share rules satisfying normalization, own-coordinate monotonicity, and efficiency (sum of credit shares equals agent utility). They establish PPAD membership under efficient evaluation and polynomially bounded $L/\eps$, and leave polynomial-time computation unresolved. Their work identifies two natural classes with opposite informational interactions.
\begin{enumerate}
\item \emph{Submodular utilities and cross-monotone shares.} Diminishing marginal returns capture overlapping informational gains: measurements covering the same phenomenon can become less useful after related measurements have arrived. Examples include information-gain objectives for noisy temperature sensors and attribute selection under suitable conditional-independence assumptions~\cite{KrauseGuestrin2005}. Overlapping patient cohorts or different noisy signals of the same population under Gaussian inference motivate the same modeling choice. With Shapley sharing, submodularity makes a donor's credit nonincreasing in every other donor's contribution. Akrami et al.\ give a convergent local-search algorithm and prove CLS membership when $L/\eps$ is polynomially bounded, but do not obtain a polynomial iteration bound.
\item \emph{Supermodular utilities and complementary shares.} Synergies arise when different datasets become more useful together. Vertical federated learning combines different features about common users~\cite{Liu2024}; multimodal medical prediction combines histology and genomic biomarkers~\cite{Mobadersany2018}. Such applications motivate synergies between datasets and suggest complementary utilities. To this end, we study lattice-supermodular utilities, where Shapley shares have the opposite cross effect: another donor's contribution can only increase a donor's credit. The computational complexity of reciprocal and core-stable data exchange in this setting is also left as an open problem in~\cite{Akrami2025}.
\end{enumerate}

\paragraph{Our contributions.}
We resolve approximate computation in these two classes. Our algorithm applies to all cross-monotone shares, including Shapley shares under submodular utilities, while our hardness result holds already for Shapley shares.
Throughout, approximation is additive, utilities lie in $[0,1]$, and vanish at zero. The hardness instances have polynomial descriptions and constant Lipschitz bounds.
\begin{enumerate}
\item \emph{An FPTAS for diminishing returns.}
We compute an exactly core-stable and $\eps$-reciprocal exchange for any cross-monotone shares.
The algorithm uses a number of share queries polynomial in the number of agents, $1/\eps$, and $L$ (an own-coordinate Lipschitz bound on the shares). For polynomially bounded numerical $L$ and efficient approximate-share access, this is an FPTAS. It applies to monotone submodular utilities with approximate Shapley oracles, strictly improving the prior local-search guarantee to polynomial-time approximation.
\item \emph{PPAD hardness for complementary datasets.} With monotone lattice-supermodular utilities and Shapley sharing, the approximate exchange problem is PPAD-hard at inverse-polynomial accuracy, even when every utility and share is $6$-Lipschitz and shares are exactly evaluable in polynomial time. Thus the difficulty is finding the exchange, rather than evaluating Shapley values. Whereas the discussion in Akrami et al.~\cite{Akrami2025} identifies analogous algorithmic obstacles in both classes, we show that the two classes have different computational complexity.
\end{enumerate}

\subsection{Technical Overview}
\label{sec:overview}

In this section, we highlight the main technical contributions of our work. We begin by reviewing the algorithm of Akrami et al.~\cite{Akrami2025}, then use its structure and limitations to explain our algorithmic advance and the hardness of complementary datasets.

\paragraph{The algorithm of Akrami et al.}
Starting from full sharing, their algorithm maintains approximate core stability by keeping the exchange graph $G(x,\alpha)$ acyclic, where an edge $i\to j$ indicates that $i\ne j$ and $x_{ij}<1-\alpha$. In any deviating coalition, the earliest agent in a topological order already receives almost all the data available within that coalition, so an internal exchange can improve its utility by at most $nL\alpha$. This gives approximate core stability, and exact core stability when $\alpha=0$. The algorithm then balances surpluses $\Delta_i(x)=\sum_j c_{ij}(x)-u_i(x)$, where $c_{ij}(x)$ is the credit earned by agent $i$ for the data it shares with $j$. It decreases data shared from high-surplus to low-surplus agents or increases sharing in the opposite direction. Changing one contribution can affect several agents' credits. With cross-monotone shares, coordinated decreases lower high surpluses without lowering low surpluses. Increases also lower high surpluses, but may lower the surpluses of low-surplus agents. Their lexicographic surplus potential establishes convergence, but does not give a polynomial iteration bound.

\paragraph{Our main algorithmic contribution.}
\emph{We show that the decrease operation alone can suffice: starting from full sharing, our algorithm reaches approximate reciprocity without ever increasing a data fraction.} Instead of directly balancing surpluses, we maintain a circulation of credit \emph{targets} $f_{ij}\le c_{ij}(x)$. These targets satisfy flow conservation by construction. We alternately increase the targets while keeping the exchange fixed and reduce the amounts of data shared to bring actual credits closer to their targets---as soon as $f$ equals $c$, we get reciprocity. Throughout, our data reductions preserve the acyclicity of the exchange graph, guaranteeing core stability.

\paragraph{The structural distinction between substitutes and complements.}
Fix an agent $j$ and a set of agents sharing data with it. Under submodular utilities, Shapley shares exhibit cross-monotonicity, i.e., the credit share of one agent can only decrease when another agent increases her contribution to the same recipient. Under cross-monotone shares, we can bring the credits of any set of agents down to any specified nonnegative targets below their current credits, without lowering the credits of agents outside this set. The reason is simple: when one agent shares less data with $j$, only its own credit at $j$ can fall; everyone else's credit at $j$ can only rise. Any excess credit created for another agent in the selected set can be removed by reducing that agent's contribution. With complementary datasets, however, sharing less data can make other agents' data (outside of the selected set) less valuable as well, reducing their credits. Bringing one agent closer to its target can therefore create a credit shortfall for another. Our hardness reduction exploits these dependencies: we design utilities so that core stability and reciprocity force the shared data fractions to encode a solution to a PPAD-hard fixed-point problem.
In the rest of this section, we give an overview of the techniques behind our two main results, with complete details deferred to Sections~\ref{sec:alg} and~\ref{sec:complexity}.

\paragraph{Overview of our FPTAS.}
Our algorithm maintains a nonnegative \emph{benchmark circulation} $f$ below the actual shares:
\begin{equation}
f_{ij}\le c_{ij}(x),\qquad
\sum_j f_{ij}=\sum_j f_{ji}\quad(i\in[n]).
\label{eq:overview-circulation}
\end{equation}
If actual credits equal this circulation, i.e., $f_{ij} = c_{ij}(x)$, reciprocity follows immediately. We begin with $x=\one$ and $f=0$, then alternate target increases and data reductions. The overall goal is to bring $f$ and $c$ close to each other while also maintaining core stability.

\emph{Get $f$ as close as possible to $c$.} Define the \emph{slack graph} on unresolved agents by placing an edge $i\to j$ whenever $f_{ij}<c_{ij}$, meaning that this target has room to increase. Along a directed cycle, we can increase all targets by the same amount while preserving flow conservation. We augment until one target reaches its credit, removing that edge from the slack graph. Repeating this operation eliminates all cycles, bringing $f$ closer to $c$ until no further cycle augmentation is possible.

\emph{Get $c$ as close as possible to $f$.} We progressively \emph{resolve} agents by freezing the data they receive and their incident targets; all other agents are \emph{unresolved}. Initially, every agent is unresolved. Choose a source $v$ of the slack graph on unresolved agents. Its incoming credits are already close to their targets: there is no incoming slack from unresolved donors, and contributions from resolved donors were repaired earlier. We can therefore resolve $v$. The remaining difficulty concerns outgoing credits. Keeping $f$ fixed, we carefully decrease contributions from all resolved donors to unresolved recipients, bringing their earned credits down toward their targets. These controlled reductions follow the same basic idea as the contribution-decrease operation of Akrami et al.~\cite{Akrami2025}. Each reduction keeps the donor's own credit above its target, while cross-monotonicity ensures that other donors' credits do not fall. Any new slack cycles are handled in the next round.

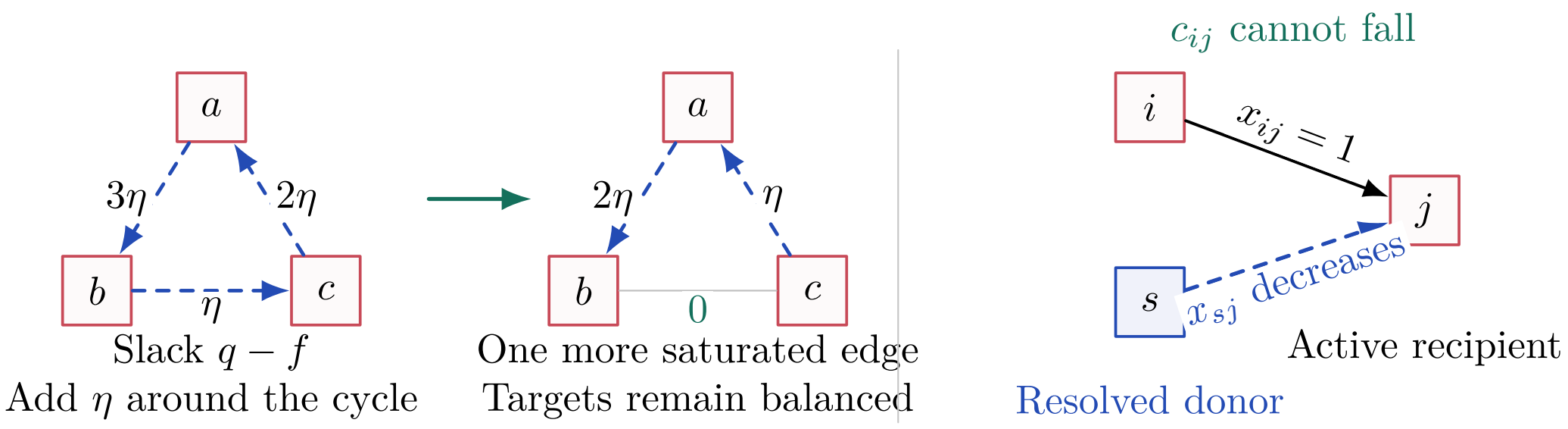
\begin{figure}[t]
\centering
\begin{tikzpicture}[font=\small]
\begin{scope}
\node[agent] (a) at (0,1) {$a$};
\node[agent] (b) at (-1,-.6) {$b$};
\node[agent] (c) at (1,-.6) {$c$};
\draw[slackedge] (a)--node[elabel,left] {$3\eta$}(b);
\draw[slackedge] (b)--node[elabel,below] {$\eta$}(c);
\draw[slackedge] (c)--node[elabel,right] {$2\eta$}(a);
\node[align=center] at (0,-1.35) {Slack $q-f$\\Add $\eta$ around the cycle};
\end{scope}
\draw[targetedge] (1.9,.2)--(2.8,.2);
\begin{scope}[xshift=4.25cm]
\node[agent] (aa) at (0,1) {$a$};
\node[agent] (bb) at (-1,-.6) {$b$};
\node[agent] (cc) at (1,-.6) {$c$};
\draw[slackedge] (aa)--node[elabel,left] {$2\eta$}(bb);
\draw[gray!50] (bb)--node[elabel,below,text=datateal] {$0$}(cc);
\draw[slackedge] (cc)--node[elabel,right] {$\eta$}(aa);
\node[align=center] at (0,-1.35) {One more saturated edge\\Targets remain balanced};
\end{scope}
\draw[gray!45] (6,-1.75)--(6,1.5);
\begin{scope}[xshift=8.2cm]
\node[agent] (i) at (0,1) {$i$};
\node[fixedagent] (s) at (0,-.7) {$s$};
\node[agent] (j) at (2.4,.1) {$j$};
\draw[dataedge] (i)--node[elabel,above,sloped] {$x_{ij}=1$}(j);
\draw[->,datablue,dashed,line width=1pt] (s)--node[elabel,below,sloped] {$x_{sj}$ decreases}(j);
\node[align=center,text=datablue] at (0,-1.55) {Resolved donor};
\node[align=center] at (2.4,-1.1) {Active recipient};
\node[align=center,text=datateal] at (1.25,1.65) {$c_{ij}$ cannot fall};
\end{scope}
\end{tikzpicture}
\caption{Left: a target-circulation augmentation preserves balance and saturates an edge. Labels are slack capacities, not data fractions. Right: a resolved donor reduces data to an active recipient. Its own share may fall, but the active donor's share cannot fall, so the circulation's surviving capacities remain feasible.}
\label{fig:algorithm-overview}
\end{figure}

Core stability follows because we only reduce contributions from resolved agents to agents that are still unresolved. Thus, in any coalition, the first agent we resolve continues to receive full data from every other member and cannot improve through an internal exchange, preventing the coalition from making all its members better off. Each round resolves at least one new agent, so the algorithm terminates after at most $n$ rounds.

\paragraph{Overview of PPAD Hardness.} Given a loop-free directed graph $G=([m],E)$ and an accuracy parameter $\delta>0$, the \emph{Directed Public-Goods Equilibrium} (\textsc{DPGE}) problem~\cite{PapadimitriouPeng2021} asks for efforts $s\in[0,1]^m$ satisfying $\left|s_i-\max\{0,1-\sum_{j:(j,i)\in E}s_j\}\right|\le\delta$ for every vertex $i$. We reduce from \textsc{DPGE} by constructing a star exchange with one leaf agent for each vertex in $G$ and one center agent: each leaf values only the center's data, while the center values data from all leaves. Each effort $s_j$ is encoded as a function of the data that leaf $j$ contributes to the center. We design the center's utility so that leaf $i$'s Shapley credit reflects both its own effort $s_i$ and the incoming total $\sum_{j:(j,i)\in E}s_j$. \emph{Complementarity makes this possible:} a multiplicative term that vanishes whenever any member of a donor group is absent divides its Shapley credit equally among the group's members. For each $j$, we include such a term involving $j$ and its outgoing neighbors in $G$, allowing the encoded effort $s_j$ of the corresponding leaf $j$ to contribute to each of their credits.

To enforce the required relation between $s_i$ and $\sum_{j:(j,i)\in E}s_j$, we use reciprocity to equate each leaf's earned credit with the utility it receives, and core stability to force that utility to its maximum (a fixed value). Specifically, the utilities are chosen so that, at a reciprocal exchange, any leaf receiving less than its maximum utility could form a blocking pair with the center through full bilateral exchange. Core stability rules this out, so reciprocity fixes each leaf's earned credit at the corresponding target. These credit constraints then enforce \textsc{DPGE}'s local condition: supply the deficit left by incoming neighbors, or zero if they already provide at least one. The argument is robust to approximation, yielding hardness at inverse-polynomial accuracy.

\subsection{Further Related Work}
\label{subsec:further-related-work}

Nonmonetary data exchange has been studied through utility balance, coalitional stability, and incentives for data collection. Bhaskara et al.~\cite{Bhaskara2024} introduce a data-exchange marketplace based on interim utility balance and study welfare maximization and core stability. Their blocking exchanges must themselves satisfy balance, whereas the core-stability notion adopted here allows arbitrary internal exchanges by a deviating coalition.
More recently, Hakim et al.~\cite{Hakim2026} study strategic data collection under contracts requiring equal quantities of data to be exchanged in both directions, obtaining efficiently computable Nash equilibria and Pareto-optimal outcomes under their assumptions. Their reciprocity concerns data quantities rather than utility contributions. Song et al.~\cite{Song2025} study a different exchange model with explicit sharing costs, establishing core existence under concave benefits and convex costs, as well as PPAD-hardness of finding a core-stable exchange. Incentives for data sharing are also central to federated learning. Blum et al.~\cite{Blum2021} study incentive-aware sample collection and its efficiency, while Donahue and Kleinberg~\cite{DonahueKleinberg2021} analyze voluntary participation and stable model-sharing coalitions among heterogeneous learners. Karimireddy et al.~\cite{KarimireddyGuoJordan2022} design accuracy-shaping mechanisms that discourage free-riding by linking the quality of the model an agent receives to its data contribution, without monetary transfers. Chaudhury et al.~\cite{RayChaudhury2022Core,RayChaudhury2024PVC} develop coalitional fairness guarantees for selecting a shared predictor, including guarantees based on the proportional veto core. Murhekar et al.~\cite{Murhekar2023Incentives,Murhekar2025Reciprocal} design budget-balanced payment mechanisms for welfare maximization and Shapley-based reciprocal fairness. These approaches complement our requirement that reciprocity be achieved through the data exchange itself. A related literature studies monetary data markets: Agarwal et al.~\cite{Agarwal2019} develop an algorithmic marketplace for training data, addressing pricing and revenue allocation; Acemoglu et al.~\cite{Acemoglu2022} show how informational externalities can depress data prices and induce excessive sharing; and Agarwal et al.~\cite{Agarwal2024Externalities} study data auctions with externalities among competing buyers.

\paragraph{Organization.} Section~\ref{sec:prelim} introduces notation and definitions. Section~\ref{sec:alg} presents the algorithm, including the full implementation of Controlled Decrease (Section~\ref{sec:controlled-decrease}), and Section~\ref{sec:complexity} proves the hardness result.
\section{Preliminaries}
\label{sec:prelim}

There are $n$ agents. An exchange is $x\in[0,1]^{n\times n}$, where $x_{ij}$ is donor $i$'s data fraction supplied to recipient $j$. Recipient $j$ has a continuous, monotone utility $u_j:[0,1]^n\to[0,1]$, with $u_j(0)=0$, depending only on its column $x_{\cdot j}$. We also write $u_j(x)$ for $u_j(x_{\cdot j})$. Donor $i$'s share $c_{ij}(x)$ at recipient $j$ likewise depends only on the column $x_{\cdot j}$, and we also write $c_{ij}(z)$ for $z\in[0,1]^n$. Shares are continuous, nonnegative, and satisfy
\begin{equation}
\begin{aligned}
&c_{ij}(z)=0\text{ if }z_i=0 &&\text{(normalization)},\\
&c_{ij}\text{ is nondecreasing in }z_i &&\text{(monotonicity)},\\
&\textstyle\sum_ic_{ij}(z)=u_j(z) &&\text{(efficiency)}.
\end{aligned}
\label{eq:admissible}
\end{equation}

\begin{definition}
An exchange is $\eps$-reciprocal if $|\Delta_i(x)|\le\eps$ for all $i$, where
\begin{equation}
\Delta_i(x)=\sum_jc_{ij}(x)-u_i(x)
=\sum_jc_{ij}(x)-\sum_jc_{ji}(x).
\label{eq:surplus}
\end{equation}
It is $\eps$-core-stable if no nonempty coalition $T$ has an internal exchange improving every member by more than $\eps$. Equivalently, by monotonicity, every nonempty $T$ contains some $i$ with $u_i(\one_T)\le u_i(x)+\eps$, where $\one_T$ is full data from the donors in $T$ and zero data from outsiders. Exact core stability takes $\eps=0$.
\end{definition}

Shares are \emph{cross-monotone} when $c_{ij}(z)$ is nonincreasing in every coordinate $z_k$, $k\ne i$. We assume a known rational $L\ge1$ such that every share is $L$-Lipschitz in its own coordinate, with other coordinates fixed. A share query returns a rational estimate within a requested additive tolerance $\tau$, with polynomial bit length. We count oracle calls separately; the overall FPTAS requires each call to take time polynomial in the input size and $1/\tau$. All guarantees below allow deterministic, adversarial errors within that tolerance.

For $z\in[0,1]^n$ and $S\subseteq[n]$, let $z[S]$ denote the bundle with $z[S]_k=z_k$ for $k\in S$ and $z[S]_k=0$ for $k\notin S$. The genuine Shapley share is
\begin{equation}
c^{\mathrm{Sh}}_{ij}(z)=\mathbb E_{\pi}
\bigl[u_j(z[P_i^\pi\cup\{i\}])-u_j(z[P_i^\pi])\bigr],
\label{eq:shapley}
\end{equation}
where $\pi$ is a uniformly random permutation of the donors $[n]$ and $P_i^\pi$ is the set of donors preceding $i$ in $\pi$. For $z,w\in[0,1]^n$, let $z\vee w$ and $z\wedge w$ denote the coordinatewise maximum and minimum. A utility $u$ is \emph{lattice-submodular} if
\[
u(z)+u(w)\ge u(z\vee w)+u(z\wedge w)\qquad\text{for all } z,w\in[0,1]^n,
\]
and \emph{lattice-supermodular} if the reverse inequality holds for all $z,w$.

\begin{proposition}\label{prop:shapley-cross}
For monotone lattice-submodular utilities, Shapley shares satisfy~\eqref{eq:admissible} and cross-monotonicity. If $u_j$ is coordinatewise $L$-Lipschitz, its shares are own-coordinate $L$-Lipschitz. For monotone lattice-supermodular utilities, the shares instead are nondecreasing in the other coordinates.
\end{proposition}
\begin{proof}
Each permutation marginal is nonnegative, zero when the donor supplies zero, nondecreasing in the donor's own coordinate, and own-coordinate $L$-Lipschitz when the utility is. Efficiency follows by telescoping along a permutation. If another donor precedes $i$, decreasing differences imply that increasing its data decreases $i$'s marginal; if it follows $i$, the marginal is unchanged. Averaging proves cross-monotonicity. Increasing differences give the reversed cross effect for supermodular utilities.
\end{proof}

\section{The Algorithm}\label{sec:alg}

\subsection{The circulation algorithm}
Our algorithm operates with actual credits $c_{ij}(x)$, and targets $f_{ij}$ on every edge $i\to j$. Throughout the algorithm,
\begin{equation}
0\le f_{ij}\le c_{ij}(x),\qquad
\sum_j f_{ij}=\sum_j f_{ji}\quad(i\in[n]).
\label{eq:benchmark}
\end{equation}
We set $f_{ii}=0$. Thus $f$ is a circulation on all agents, including agents whose targets have already been fixed. Initially, $x=\one$ and $f=0$.

An agent is declared \emph{resolved} after its received-data column and incident targets have been fixed and its outgoing credits have been brought close to their targets. Let $S$ be the resolved set and $R$ the active set. While a selected source is being processed, it is temporarily outside both sets. Only resolved donors and this selected source may reduce data, and only to the remaining active recipients. Initially $S=\varnothing$ and $R=[n]$.

\paragraph{Phase~1: increasing targets to match credits via cycle augmentation.} In Phase~1, we keep the exchange $x$ and the credits $c_{ij}(x)$ fixed and move the targets as close as possible to the credits through cycle augmentations. To accommodate approximate share queries and keep all arithmetic on rational numbers of polynomial bit length, we introduce the following parameters:
\begin{equation}
\eta=\frac{\eps}{2n},\qquad
\tau=\frac{\eps}{32n^3},\qquad
\Gamma=\frac{3\eps}{4n}.
\label{eq:parameters}
\end{equation}
Every share query has additive error at most $\tau$.
For each ordered pair of distinct active agents, we maintain a lower credit capacity $q_{ij}$, initially zero. Before the first round and after each completed resolution, query $\widehat c_{ij}$ and update
\begin{equation}
b_{ij}=\eta\max\left\{0,\left\lfloor
\frac{\widehat c_{ij}-\tau}{\eta}\right\rfloor\right\},
\qquad q_{ij}\gets\max\{q_{ij},b_{ij}\}.
\label{eq:refresh}
\end{equation}
Subtracting $\tau$ before rounding makes $b_{ij}$ a lower bound. The \emph{slack graph} $H$ has the active agents $R$ as its vertices and an edge $i\to j$ whenever $i\ne j$ and $f_{ij}<q_{ij}$. To augment a directed cycle $C$, use
\begin{equation}
\lambda=\min_{(i,j)\in C}(q_{ij}-f_{ij}),\qquad
f_{ij}\gets f_{ij}+\lambda\quad((i,j)\in C).
\label{eq:augment}
\end{equation}
Neither $x$ nor $q$ changes during these augmentations. The graph contains only edges on which targets can increase.

\paragraph{Phase~2: decreasing credits to match targets.}
The goal is now to bring $c$ as close as possible to $f$ by decreasing data sent from resolved agents to active agents. However, decreasing one resolved agent's contribution to an active recipient can increase another resolved agent's credit at that recipient. This moves the other outgoing edge farther above its fixed target. We therefore need a coordinated decrease of these contributions. Akrami et al.~\cite[Algorithm~2]{Akrami2025} provide a decreasing-flow procedure that addresses this interaction. We use it through the following interface; Section~\ref{sec:controlled-decrease} gives the procedure and proves correctness and termination.

\begin{definition}[Controlled Decrease]
\label{def:controlled-decrease}
\emph{Input.} An exchange $x$ reached by Algorithm~\ref{alg:circulation}, fixed targets $0\le f\le c(x)$, an active recipient $j$, and a donor set $A$ with $j\notin A$. The accuracy parameters are those in~\eqref{eq:parameters}.

\emph{Output.} The operation \textsc{ControlledDecrease}$(j,A)$ returns an exchange $x'$ obtained by decreasing only the fractions $x_{ij}$ with $i\in A$, such that
\begin{equation}
0\le c_{ij}(x')-f_{ij}\le\Gamma\quad(i\in A).
\label{eq:controlled-bound}
\end{equation}
The targets remain feasible, $f\le c(x')$. For every donor $k\notin A$, its credit at $j$ does not decrease: $c_{kj}(x')\ge c_{kj}(x)$.

\emph{Running time.} A call uses a number of share queries at tolerance $\tau$ that is polynomial in $n$, $L$, and $1/\eps$, with polynomial arithmetic overhead.
\end{definition}

Phase~2 completes all outgoing-credit adjustments for the current source before declaring it resolved. For a selected source $v$, set $A=S\cup\{v\}$ and run the subroutine to completion at every remaining active recipient. This makes every outgoing non-self edge from $A$ almost saturated, meaning $0\le c_{ij}-f_{ij}\le\Gamma$. It is important to include the old resolved donors: another donor's decrease can raise their credits again.

After this pass, refresh the remaining active capacities and continue with any new source. Thus one Phase~2 can resolve several agents. Return to Phase~1 only when active agents remain but the refreshed slack graph has no source. In that case it contains a directed cycle.

\begin{algorithm}[t]
\caption{The circulation algorithm with repeated source resolution}
\label{alg:circulation}
\begin{algorithmic}[1]
\State $x\gets\one$.
\If{$n=1$}
  \State \Return $x$ and order $(1)$.
\EndIf
\State Set parameters~\eqref{eq:parameters}; $f,q\gets0$.
\State $S\gets\varnothing$; $R\gets[n]$; omit self-loops in $f,q$.
\State Refresh $q$ by~\eqref{eq:refresh}; form the slack graph $H$.
\While{$R\ne\varnothing$}
  \Statex \hspace{\algorithmicindent}\textbf{Phase 1: bring $f$ closer to $c$.}
  \While{$H$ contains a directed cycle $C$}
    \State Augment $f$ on $C$ by~\eqref{eq:augment}; update $H$.
  \EndWhile
  \Statex \hspace{\algorithmicindent}\textbf{Phase 2: bring $c$ closer to $f$.}
  \While{$H$ has a source $v$}
    \State Fix $x_{\cdot v}$ and all targets entering or leaving $v$.
    \State $A\gets S\cup\{v\}$; $R\gets R\setminus\{v\}$.
    \For{each $j\in R$}
      \State Run \textsc{ControlledDecrease}$(j,A)$ to completion.
    \EndFor
    \State $S\gets A$. \Comment{The outgoing credits are now close.}
    \State Refresh $q$ on $R$ by~\eqref{eq:refresh}; rebuild $H$.
  \EndWhile
\EndWhile
\State \Return $x$ and the resolution order.
\end{algorithmic}
\end{algorithm}

\subsection{Implementing Controlled Decrease with the Original Decrease Routine}
\label{sec:controlled-decrease}

We implement the interface in Definition~\ref{def:controlled-decrease} by retaining the relative-floor decrease loop of Akrami et al.~\cite[Algorithm~2 and Lemma~11]{Akrami2025}. Its inputs are a recipient $j$ and a donor set $A$ not containing $j$. In the main algorithm, $A$ consists of the resolved donors and the source currently being processed. An eligibility test selects the donors to which the original loop can safely be applied. We give the full implementation because the one-sided guarantees must hold with approximate share queries.

Throughout this subsection, $n\ge2$. Using the oracle tolerance $\tau=\eps/(32n^3)$ from the main algorithm, set
\[
\alpha=16\tau=\frac{\eps}{2n^3}.
\]
All share estimates have additive error at most $\tau$. Data fractions lie on the common dyadic grid of mesh
\[
\rho=2^{-K},\qquad
K=\left\lceil\log_2\frac{L}{\tau}\right\rceil,
\]
so $L\rho\le\tau$. The integer $K$ can be computed by exact comparisons with powers of two. Full sharing lies on this grid, and every update below preserves it.

\subsubsection{The original relative-floor loop and the eligibility wrapper}

The original routine fixes a donor set and one recipient. At the start of that call it assigns each donor the surplus floor
\[
\delta_i=\Delta_i(x^{\rm start})-\alpha.
\]
While a donor's surplus exceeds its floor by more than $\alpha/2$, the routine lowers that donor's data to the recipient until the floor is reached, or until no data remain. It then updates the surpluses and repeats. The original implementation realizes floor hitting by binary search, and its Lemma~11 bounds the number of updates using the data removed in each update.

For a fixed recipient $j$ and any donor $i\ne j$, write
\[
\Delta_i(x)=c_{ij}(x)+B_i,
\qquad
B_i=\sum_{k\ne j}c_{ik}(x)-u_i(x).
\]
Only column $j$ changes during the call, so $B_i$ is constant. Thus the original surplus floor is exactly the credit floor $c_{ij}(x^{\rm start})-\alpha$. In particular, its tests can be performed using credits alone, without computing $B_i$.

Our wrapper applies this same loop only to donors whose estimated credit exceeds $f_{ij}+2\alpha$. If $a_i$ denotes the estimate used for this selection, the relative credit floor is
\[
p_i=a_i-\alpha>f_{ij}+\alpha.
\]
This margin protects the circulation target even though the original floor is chosen relative to the start of the call. With exact estimates, it is precisely the original floor choice. With approximate estimates, the chosen floor differs by at most $\tau$. Since $p_i>0$ and a zero donation has zero credit, the original routine's zero-data fallback is unnecessary in these calls.

\begin{algorithm}[t]
\caption{$\textsc{ControlledDecrease}(j,A)$: eligibility wrapper}
\label{alg:controlled-decrease}
\begin{algorithmic}[1]
\Require Current $x$, fixed targets $f$, active recipient $j$, selected donors $A$
\Loop
  \State Query $a_i$ estimating $c_{ij}(x)$ for every $i\in A$.
  \State $D\gets\{i\in A:a_i>f_{ij}+2\alpha\}$.
  \If{$D=\varnothing$}
    \State \Return $x$.
  \EndIf
  \State Apply Algorithm~\ref{alg:original-decrease} to $(x,j,D,(a_i)_{i\in D})$.
\EndLoop
\end{algorithmic}
\end{algorithm}

Algorithm~\ref{alg:original-decrease} retains the original relative-floor rule, trigger, and repeated donor updates. Its binary-search queries concern hypothetical bundles: the actual exchange changes only at line~\ref{line:commit-decrease}. In the pseudocode, $\gamma_i(z)$ is donor $i$'s share at recipient $j$ when only $x_{ij}$ is replaced by $z$ in the current exchange.

\begin{algorithm}[t]
\caption{Original relative-floor decrease loop with approximate queries}
\label{alg:original-decrease}
\begin{algorithmic}[1]
\Require Current $x$, recipient $j$, selected donors $D$, their starting estimates $a_i$
\State $p_i\gets a_i-\alpha$ for every $i\in D$.
\Loop
  \State Query $\widehat c_{ij}$ for every $i\in D$.
  \If{$\widehat c_{ij}\le p_i+\alpha/2$ for every $i\in D$}
    \State \Return $x$.
  \EndIf
  \State Choose $i\in D$ with $\widehat c_{ij}>p_i+\alpha/2$.
  \State $\ell\gets0$; $r\gets x_{ij}/\rho$.
    \Comment{Integer grid indices}
  \While{$r-\ell>1$}
    \State $m\gets\lfloor(\ell+r)/2\rfloor$.
    \State Query $\widehat\gamma_i(m\rho)$ with error at most $\tau$.
    \If{$\widehat\gamma_i(m\rho)\ge p_i+\tau$}
      \State $r\gets m$.
    \Else
      \State $\ell\gets m$.
    \EndIf
  \EndWhile
  \State $x_{ij}\gets r\rho$.
    \label{line:commit-decrease}
\EndLoop
\end{algorithmic}
\end{algorithm}

After a call returns, other donors in $A$ may now exceed the eligibility threshold because cross-monotonicity can raise their credits. The wrapper therefore queries all of $A$ again and selects a new set $D$.

\subsubsection{One-sided floor hitting and termination}

\begin{lemma}[One decrease]
\label{lem:one-decrease}
Every committed update in Algorithm~\ref{alg:original-decrease} leaves the selected credit in $[p_i,p_i+3\tau]$, decreases that credit by more than $\alpha/4$, and removes more than $\alpha/(4L)$ data. It uses at most $K$ binary-search queries. Every other donor's credit at $j$ weakly increases.
\end{lemma}
\begin{proof}
At the start of the search, the trigger and oracle accuracy give
\[
\gamma_i(x_{ij})>p_i+\alpha/2-\tau>p_i,
\qquad \gamma_i(0)=0<p_i.
\]
The search maintains
\[
\gamma_i(r\rho)\ge p_i,
\qquad \gamma_i(\ell\rho)<p_i+2\tau.
\]
Indeed, an estimate at least $p_i+\tau$ certifies that its true value is at least $p_i$. An estimate below that threshold certifies a true value below $p_i+2\tau$. These are the only facts required of a noisy comparison; the oracle errors need not have a consistent sign.

The initial index interval has length at most $2^K$, so at most $K$ comparisons leave adjacent endpoints. Own-coordinate Lipschitz continuity then gives
\[
p_i\le \gamma_i(r\rho)
\le \gamma_i(\ell\rho)+L\rho
<p_i+3\tau.
\]
The credit loss from the committed update is therefore greater than
\[
(\alpha/2-\tau)-3\tau
=\alpha/2-4\tau=\alpha/4.
\]
An $L$-Lipschitz credit cannot lose this much unless the donation decreases by more than $\alpha/(4L)$. Since the update only reduces one donation, cross-monotonicity makes all other donors' credits weakly increase.
\end{proof}

\begin{lemma}[Correctness and termination of Controlled Decrease]
\label{lem:controlled-decrease}
For every input in Definition~\ref{def:controlled-decrease}, Algorithm~\ref{alg:controlled-decrease} terminates and returns an output satisfying that definition, within its stated running time. Across all calls made by Algorithm~\ref{alg:circulation}, the total number of queries is bounded by~\eqref{eq:query-cap}.
\end{lemma}
\begin{proof}
Every selected floor satisfies $p_i>f_{ij}+\alpha$. Lemma~\ref{lem:one-decrease} therefore preserves the selected donor's target lower bound. All other credits weakly increase during that update, so every initially valid target lower bound in the column remains valid. Only donors in $A$ reduce data; consequently, every donor outside $A$ has a nondecreasing credit throughout the wrapper.

Each nonempty call to Algorithm~\ref{alg:original-decrease} performs at least one update. To see this, note that its starting estimate $a_i$ and the estimate in its first scan differ by at most $2\tau$, since the exchange has not changed. Thus every selected donor initially satisfies
\[
\widehat c_{ij}-p_i\ge\alpha-2\tau>\alpha/2.
\]
Every update removes more than $\alpha/(4L)$ from one data fraction, and no fraction ever increases. Hence all inner calls and the wrapper terminate. The same argument holds across all rounds of the main algorithm: summing over the $n(n-1)$ non-self coordinates gives the global update bound
\[
T=n(n-1)\left\lceil\frac{4L}{\alpha}\right\rceil.
\]

When the wrapper returns, its last scan has $a_i\le f_{ij}+2\alpha$ for every $i\in A$. Oracle accuracy and feasibility imply
\[
0\le c_{ij}(x)-f_{ij}\le2\alpha+\tau
=\frac{33\eps}{32n^3}
\le\frac{3\eps}{4n}=\Gamma,
\qquad i\in A,
\]
where the last inequality uses $n\ge2$. All other columns are unchanged, so this proves the output guarantees. The per-call and aggregate running-time bounds follow from the counts below.
\end{proof}

\subsubsection{Query cap and rational encodings}

\paragraph{A single call.}
Let $m=|A|$ and $B=\lceil4L/\alpha\rceil$. Each of the $m$ affected data fractions can decrease at most $B$ times, so one call makes at most $mB$ committed updates. Every nonempty inner call performs an update. The scans that select donors, the final inner scans, and the wrapper's nonempty-selection scans therefore cost at most $3m$ queries per committed update in total. Binary search adds at most $K$ queries per update, and the final empty-selection scan costs $m$. Thus the number of queries in one call is at most
\[
Q_j\le mB(3m+K)+m
=O\!\left(\frac{n^4L}{\eps}
\left[n+\log\frac{nL}{\eps}\right]\right).
\]
Here we used $m\le n$, $\alpha=\eps/(2n^3)$, and $K=O(\log(nL/\eps))$. If $A$ is empty, the procedure returns without a share query.

\paragraph{All calls in the main algorithm.}
For completeness, let $U$ be the actual number of committed updates over the entire main algorithm, and let $M$ be the number of nonempty calls to the original relative-floor loop. The preceding proof gives $M\le U\le T$. Its scans that lead to updates use at most $nU$ queries, and its final scans use at most $nM$. The wrapper's scans that select a nonempty $D$ use at most another $nM$ queries. Binary searches contribute at most $KU$ queries.

There is one additional, empty-selection scan per call to $\textsc{ControlledDecrease}$. The main algorithm calls the wrapper once for each remaining active recipient while processing a selected source. There is one capacity refresh on the initial active set and one after each completed resolution. These refreshes, together with the final wrapper scans, use fewer than $2n^3$ queries. Thus a deterministic cap for the entire main algorithm is
\begin{equation}
T=n(n-1)\left\lceil\frac{4L}{\alpha}\right\rceil,\qquad
Q =T(3n+K)+2n^3.
\label{eq:query-cap}
\end{equation}
In particular,
\[
Q=O\!\left(\frac{n^5L}{\eps}
\left[n+\log{\frac{n^3L}{\eps}}\right]+n^3\right).
\]
The binary-search endpoints are integer grid indices with $O(K)$ bits, and every data fraction remains a multiple of $\rho$. The stored estimates and relative floors have polynomial bit length by the oracle assumption. Hence the implementation has polynomial arithmetic overhead. If the approximate-share oracle is randomized, the algorithm can enforce the same cap; when all estimates satisfy their error bounds, no cap is reached prematurely.

\paragraph{Why active donors cannot all decrease.}
Consider three agents with $u_0=0$, $u_1=\max\{x_{01},x_{21}\}$, and $u_2=\max\{x_{02},x_{12}\}$, using genuine Shapley shares. These utilities are monotone and lattice-submodular. At full sharing, each nonzero share equals $1/2$, and $f_{12}=f_{21}=1/2$, with every other target zero, is a feasible circulation. Agent $0$ is the unique source of the slack graph formed from the actual credits. Lowering $x_{01}$ and $x_{02}$ to zero preserves core stability and all target lower bounds, but increases both $c_{12}$ and $c_{21}$ to one. The two remaining active agents now form a slack cycle.

If both active donors then reduce their mutual contributions to $x_{12}=x_{21}=1/2$, their credits equal the circulation exactly and the exchange is reciprocal. Nevertheless, coalition $\{1,2\}$ blocks: full bilateral sharing raises both utilities from $1/2$ to one. Thus decreasing every active donor's outgoing credits toward its target can destroy exact core stability. The example uses actual-credit capacities to display the mechanism; the strict blocking improvement persists under sufficiently small perturbations. The algorithm instead refreshes its certified capacities after each source pass and augments any remaining cycles once no source is available.

\subsection{Invariants and correctness}

The invariant is stated between completed resolutions. A selected source is placed next in the resolution order while its outgoing contributions are being adjusted.

\begin{lemma}[Resolution invariant]
\label{lem:invariant}
After the initial capacity refresh, after each completed resolution and refresh, and throughout Phase~1:
\begin{enumerate}
\item $f$ is a nonnegative circulation on all agents. Every target incident to a resolved agent is fixed.
\item Active agents share full data with one another. For distinct active $i,j$, $f_{ij}\le q_{ij}\le c_{ij}$ and $c_{ij}-q_{ij}\le\Gamma$.
\item For a resolved donor $i$ and active recipient $j$, $0\le c_{ij}-f_{ij}\le\Gamma$.
\item At every resolved recipient $j$, the received data are fixed and $0\le c_{ij}-f_{ij}\le\Gamma$ for every $i\ne j$.
\end{enumerate}
Every source pass in Phase~2 restores these statements with one additional resolved agent.
\end{lemma}
\begin{proof}
Initially $S$ is empty, all data fractions equal one, and $f=0$. The first capacity refresh establishes (2); the other statements are immediate. Cycle augmentation preserves balance and respects $q$. It changes neither data nor targets incident to $S$, so it preserves all four statements.

Now let $v$ be a source of the refreshed slack graph. For every other active donor $i$, the absence of $i\to v$ gives $f_{iv}=q_{iv}$, hence $c_{iv}-f_{iv}\le\Gamma$ by (2). Incoming credits from resolved donors satisfy the same bound by (3). Fixing $v$'s column and incident targets therefore gives (4) for $v$. This argument only needs $v$ to be a source; cycles elsewhere in the graph do not matter.

Remove $v$ from the active set and use $A=S\cup\{v\}$ as the donors in the decrease pass. At each remaining recipient $j$, old resolved donors have feasible targets by (3), and $v$ has a feasible target by $f_{vj}\le q_{vj}\le c_{vj}$. Lemma~\ref{lem:controlled-decrease} brings every credit from $A$ at $j$ within $\Gamma$ of its target. A call at another recipient changes a different column, so it cannot undo this bound. Thus, after all calls, every outgoing edge from $A$ to the remaining active agents satisfies (3).

These decreases change no fixed recipient column and no data between remaining active agents. Credits of these active donors cannot decrease, so the old capacities remain valid lower bounds. Refreshing them restores the error bound in (2). Targets have stayed fixed throughout the pass, preserving (1). We can now declare $v$ resolved and continue with another source, if one exists.
\end{proof}

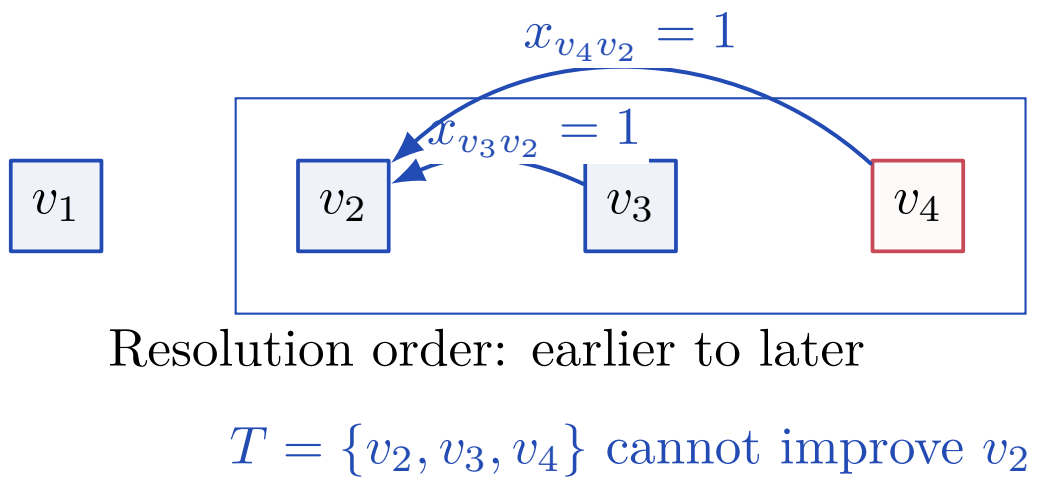
\begin{figure}[t]
\centering
\begin{tikzpicture}[font=\small,x=.95cm,y=.85cm]
\node[fixedagent] (v1) at (0,0) {$v_1$};
\node[fixedagent] (v2) at (2,0) {$v_2$};
\node[fixedagent] (v3) at (4,0) {$v_3$};
\node[agent] (v4) at (6,0) {$v_4$};
\begin{scope}[on background layer]
\node[draw=datablue,fit=(v2)(v3)(v4),inner sep=4mm] {};
\end{scope}
\draw[dataedge,datablue] (v3) to[bend right=25]
  node[elabel,above,pos=.25] {$x_{v_3v_2}=1$}(v2);
\draw[dataedge,datablue] (v4) to[bend right=42]
  node[elabel,above] {$x_{v_4v_2}=1$}(v2);
\node[align=center] at (3,-1.1)
  {Resolution order: earlier to later};
\node[align=center,text=datablue] at (4,-1.9)
  {$T=\{v_2,v_3,v_4\}$ cannot improve $v_2$};
\end{tikzpicture}
\caption{In any coalition, the earliest-resolved member already receives full data from all other members. The blue rectangle marks an example coalition. Only data sent from earlier to later agents can decrease.}
\label{fig:core-order}
\end{figure}

\begin{lemma}[Core stability]
\label{lem:core}
The exchange is exactly core-stable throughout the algorithm.
\end{lemma}
\begin{proof}
Order the resolved agents by resolution time, followed by the source currently being processed, if any, and then the remaining active agents in any order. A contribution can fall below one only from an earlier agent to a later one. Every later agent still sends full data to every earlier agent. Equivalently, the missing-data graph $G(x,0)$, with edges $i\to j$ when $i\ne j$ and $x_{ij}<1$, is acyclic.

Consider any nonempty coalition $T$ and its earliest member $i$ in this order. Agent $i$ already receives full data from every member of $T$, including itself. A coalition exchange cannot offer $i$ more of any member's data, and offers no data from outsiders. Its bundle is therefore coordinatewise bounded by the bundle $i$ currently receives. Monotonicity prevents $i$ from improving. Thus $T$ cannot make every member strictly better off.
\end{proof}

The missing-data graph used in this proof differs from the slack graph. The former stays acyclic throughout the algorithm. The latter may acquire new cycles when capacities are refreshed after a decrease pass. Phase~2 can still process any available source; if none remains, Phase~1 removes the cycles.

\begin{lemma}[Reciprocity]
\label{lem:reciprocity}
The final exchange is $\eps$-reciprocal.
\end{lemma}
\begin{proof}
For $n=1$, reciprocity is immediate. Otherwise, every recipient is resolved at termination. By the resolution invariant, write $c_{ij}=f_{ij}+e_{ij}$ with $0\le e_{ij}\le\Gamma$ for $i\ne j$. Because $f$ is balanced and self-credits cancel,
\[
\Delta_i=\sum_{j\ne i}e_{ij}-\sum_{j\ne i}e_{ji}.
\]
Each sum belongs to $[0,(n-1)\Gamma]$. Their difference therefore has absolute value at most $(n-1)\Gamma<\eps$.
\end{proof}

\begin{theorem}
\label{thm:fptas}
For cross-monotone shares satisfying the assumptions above and every rational $0<\eps\le1$, Algorithm~\ref{alg:circulation} returns an exactly core-stable, $\eps$-reciprocal exchange. The number of share queries at additive tolerance $\eps/(32n^3)$ is polynomial in $n$, $L$, and $1/\eps$, and the algorithm performs polynomially many graph operations and has polynomial bit overhead.

For polynomially bounded numerical $L$ and efficient approximate-share oracles, this is an FPTAS. It applies in particular to monotone submodular utilities with Shapley sharing.
\end{theorem}
\begin{proof}
For $n=1$, full sharing satisfies both requirements. Suppose $n\ge2$. We first show that the algorithm terminates. With $r$ active agents, each cycle augmentation saturates a new edge. Capacities $q$ stay fixed during Phase~1 and targets only increase, so a saturated edge cannot become unsaturated. Thus Phase~1 ends after at most $r(r-1)$ augmentations. Its final slack graph is acyclic, so a nonempty active set supplies at least one source pass in Phase~2. Every Controlled Decrease call terminates by Lemma~\ref{lem:controlled-decrease}, and each completed source pass resolves one new agent by Lemma~\ref{lem:invariant}. There are therefore exactly $n$ resolutions and at most $n$ outer rounds.

Lemmas~\ref{lem:core} and~\ref{lem:reciprocity} now give exact core stability and $\eps$-reciprocity of the returned exchange.

We next bound the total work. There are $O(n^3)$ cycle augmentations over all rounds. Detecting cycles and choosing sources take $O(n^2)$ graph work per call, giving $O(n^5)$ graph operations. Section~\ref{sec:controlled-decrease} bounds the share queries used by all Controlled Decrease calls together by
\[
O\!\left(\frac{n^5L}{\eps}
\left(n+\log\frac{nL}{\eps}\right)\right).
\]
Updating capacities uses only $O(n^3)$ additional queries. Equation~\eqref{eq:query-cap} gives a deterministic query cap $Q$ that includes these updates. All queries use tolerance $\tau=\eps/(32n^3)$. The implementation analysis also verifies polynomial bit length for the data fractions. Targets and capacities are multiples of the rational number $\eta$ in $[0,1]$, so they too have polynomial bit length.

With polynomially bounded numerical $L$ and efficient approximate-share oracles, the total time is polynomial in the input size and $1/\eps$. Proposition~\ref{prop:shapley-cross} gives the specialization to monotone submodular utilities with Shapley sharing.
\end{proof}

\section{PPAD Hardness for Complementary Datasets}
\label{sec:complexity}

We now show that the positive result of Section~\ref{sec:alg} does not extend to complementary data. For monotone lattice-supermodular utilities with genuine Shapley sharing, computing an approximately reciprocal and approximately core-stable exchange is $\PPAD$-hard (Theorem~\ref{comp:thm:hardness}), even when the shares are exactly computable in polynomial time and $6$-Lipschitz, and core stability is imposed only against two-agent coalitions. The difficulty therefore lies in finding the exchange, not in evaluating credits.

We reduce from the directed public-goods equilibrium problem (DPGE) of Papadimitriou and Peng~\cite{PapadimitriouPeng2021}. Given a DPGE instance on a graph $G$, we construct a star economy with a center and one leaf per vertex of $G$. The value of each source variable is encoded in the amount of data the corresponding leaf sends to the center, while the edges of $G$ are encoded in the complementary terms of the center's utility, which determine how Shapley sharing distributes credit among the leaves. The proof that every solution of the exchange problem encodes a DPGE solution proceeds in three steps.
\begin{enumerate}
\item Core stability and reciprocity force the credit of every leaf to a common target (Section~\ref{comp:subsec:targets}).
\item This target can only be met if every leaf sends more than a quarter of its data, which makes all complementary terms of the center's utility effective (Section~\ref{comp:subsec:activation}).
\item Once all these terms are effective, the reciprocity equation of each leaf coincides with the best-response condition of the corresponding vertex (Section~\ref{comp:subsec:decode}).
\end{enumerate}
Each step is first established for exact solutions and then quantified for approximate ones.

\subsection{The source problem}
\label{comp:subsec:source}

\begin{definition}[Directed Public-Goods Equilibrium (DPGE)]
\label{comp:def:dpge}
\emph{Given:} a loop-free directed graph $G=([n],E)$, with $n\ge2$, and a rational accuracy parameter $0<\delta\le1$.

\emph{Find:} a rational vector $s\in[0,1]^n$ such that
\begin{equation}
\left|s_i-\Bigl[1-\sum_{j:(j,i)\in E}s_j\Bigr]_+\right|
\le\delta
\quad (i\in[n]),
\label{comp:eq:source}
\end{equation}
where $[t]_+=\max\{t,0\}$. An \emph{exact equilibrium} satisfies~\eqref{comp:eq:source} with $\delta=0$.
\end{definition}

For a vector $s$, we write $S_i=\sum_{j:(j,i)\in E}s_j$ for the total signal that vertex $i$ receives from its in-neighbors. The best response $[1-S_i]_+$ supplies exactly the amount needed to raise the received signal to one, and nothing if the in-neighbors already supply at least one. For instance, if $E=\{(1,2),(1,3)\}$, the unique equilibrium is $s=(1,0,0)$: vertex~$1$ receives no signal and supplies one unit, which in turn satisfies vertices~$2$ and~$3$.

Papadimitriou and Peng~\cite[Theorem~5.1 and Lemma~5.3]{PapadimitriouPeng2021} prove that DPGE is $\PPAD$-hard for some inverse-polynomial $\delta$. Their Section~5.1 measures approximation by the distance between each action and its best response, exactly as in~\eqref{comp:eq:source}.

\subsection{The data-exchange instance}
\label{comp:subsec:construction}

\begin{figure}[t]
\centering
\begin{tikzpicture}[font=\small,
 compagent/.style={rectangle,draw=datablue,fill=datablue!5,line width=.7pt,minimum size=6mm,inner sep=1.5pt},
 compcenter/.style={compagent,draw=datateal,fill=datateal!8},
 compedge/.style={->,line width=.8pt},
 complabel/.style={fill=white,inner sep=1.3pt}]
\begin{scope}
\node[compagent] (j) at (0,1.2) {$j$};
\node[compagent] (i) at (-.9,-.15) {$i$};
\node[compagent] (k) at (.9,-.15) {$k$};
\draw[compedge] (j)--(i);
\draw[compedge] (j)--(k);
\node[align=center] at (0,-.92) {Source graph\\$T_j=\{j,i,k\}$};
\end{scope}
\draw[->,datablue,line width=1pt] (1.62,.55)--(2.42,.55);
\begin{scope}[xshift=4.25cm]
\node[compagent] (j2) at (0,1.2) {$j$};
\node[compagent] (i2) at (-1.05,-.15) {$i$};
\node[compagent] (k2) at (1.05,-.15) {$k$};
\node[compcenter] (c) at (0,-1.15) {$0$};
\draw[compedge,datablue] (j2)--node[complabel,right,pos=.33] {$a_j$}(c);
\draw[compedge,datablue] (i2)--node[complabel,right,pos=.4] {$a_i$}(c);
\draw[compedge,datablue] (k2)--node[complabel,right] {$a_k$}(c);
\draw[compedge,datateal] (c) to[bend left=22] node[complabel,left] {$b_i$}(i2);
\node[align=center] at (0,-1.98) {Exchange star};
\end{scope}
\draw[->,datablue,line width=1pt] (5.85,.55)--(6.65,.55);
\begin{scope}[xshift=8.7cm]
\node[draw=datateal,fill=datateal!5,inner sep=4pt,align=center] (p) at (0,1.1)
{$3P_j=3r(a_j)h(a_i)h(a_k)$};
\node[compagent] (pj) at (-1.2,-.25) {$j$};
\node[compagent] (pi) at (0,-.25) {$i$};
\node[compagent] (pk) at (1.2,-.25) {$k$};
\foreach \v in {pj,pi,pk}
 \draw[compedge,datateal] (p)--node[complabel,pos=.63] {$P_j$}(\v);
\node[align=center] at (0,-1.05) {Shapley credit of this term};
\end{scope}
\end{tikzpicture}
\caption{A vertex $j$ with out-neighbors $i,k$ yields a product term on $T_j=\{j,i,k\}$. Data flow only between the center and the leaves (middle); the source edges determine how the center's utility is credited (right). The term is a unanimity game on $T_j$, so each member receives scaled credit $P_j$. Only one return edge of the star is drawn.}
\label{comp:fig:star}
\end{figure}
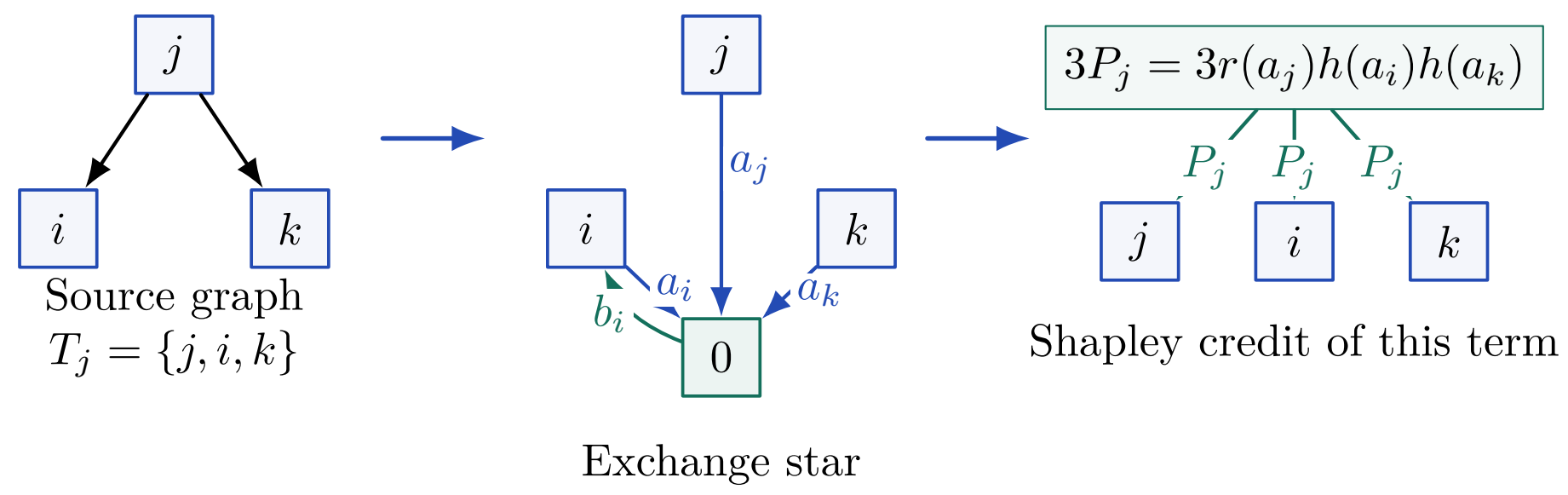

\paragraph{Agents.}
Given a DPGE instance on $G=([n],E)$, we create $N=n+1$ agents: a center $0$ and a leaf $i$ for each vertex $i\in[n]$. We write
\[
a_i=x_{i0},\qquad b_i=x_{0i}
\]
for the fraction of data that leaf $i$ sends to the center and the fraction that the center returns to it. All utilities depend only on these center--leaf coordinates; the remaining coordinates, including self-data, are ignored and may be set to zero. The utility-relevant exchange is therefore a star, regardless of the topology of $G$. The edges of $G$ enter the instance only through the center's utility (Figure~\ref{comp:fig:star}).

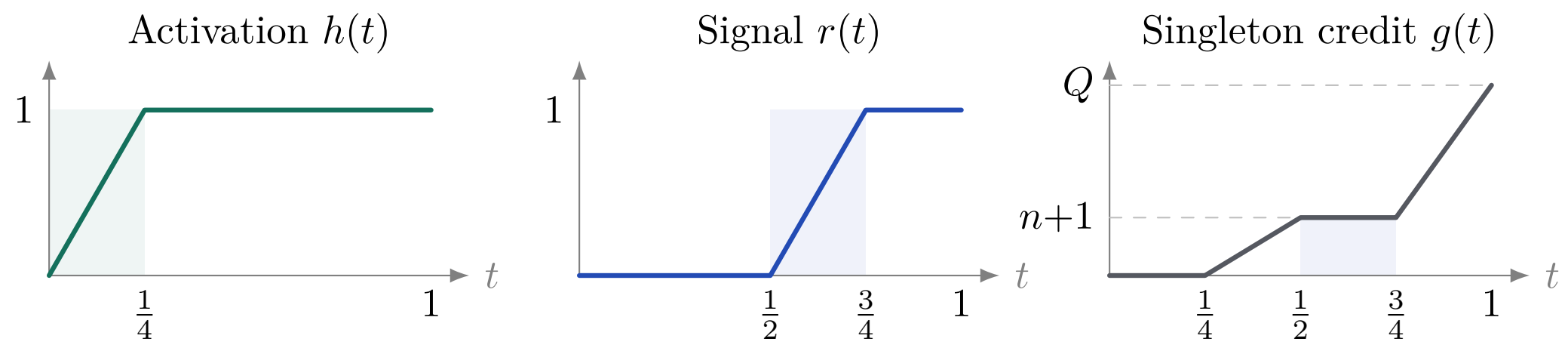
\begin{figure}[t]
\centering
\begin{tikzpicture}[font=\small,x=3.35cm,y=1.45cm]
\begin{scope}
\fill[datateal!7] (0,0) rectangle (.25,1);
\draw[->,gray] (0,0)--(1.1,0) node[right] {$t$};
\draw[->,gray] (0,0)--(0,1.3);
\draw[very thick,datateal] (0,0)--(.25,1)--(1,1);
\node[left] at (0,1) {$1$};
\node[below] at (.25,0) {$\frac{1}{4}$};
\node[below] at (1,0) {$1$};
\node[above] at (.55,1.25) {Activation $h(t)$};
\end{scope}
\begin{scope}[xshift=4.65cm]
\fill[datablue!7] (.5,0) rectangle (.75,1);
\draw[->,gray] (0,0)--(1.1,0) node[right] {$t$};
\draw[->,gray] (0,0)--(0,1.3);
\draw[very thick,datablue] (0,0)--(.5,0)--(.75,1)--(1,1);
\node[left] at (0,1) {$1$};
\node[below] at (.5,0) {$\frac{1}{2}$};
\node[below] at (.75,0) {$\frac{3}{4}$};
\node[below] at (1,0) {$1$};
\node[above] at (.55,1.25) {Signal $r(t)$};
\end{scope}
\begin{scope}[xshift=9.3cm]
\fill[datablue!7] (.5,0) rectangle (.75,.35);
\draw[->,gray] (0,0)--(1.1,0) node[right] {$t$};
\draw[->,gray] (0,0)--(0,1.3);
\draw[very thick,datagray] (0,0)--(.25,0)--(.5,.35)--(.75,.35)--(1,1.15);
\draw[dashed,gray!50] (0,.35)--(.5,.35);
\draw[dashed,gray!50] (0,1.15)--(1,1.15);
\node[left] at (0,.35) {$n{+}1$};
\node[left] at (0,1.15) {$Q$};
\node[below] at (.25,0) {$\frac{1}{4}$};
\node[below] at (.5,0) {$\frac{1}{2}$};
\node[below] at (.75,0) {$\frac{3}{4}$};
\node[below] at (1,0) {$1$};
\node[above] at (.55,1.25) {Singleton credit $g(t)$};
\end{scope}
\end{tikzpicture}
\caption{The activation, signal, and singleton functions. A leaf is active once its data fraction reaches $\frac{1}{4}$. On the shaded interval $[\frac{1}{2},\frac{3}{4}]$, the signal $r$ varies while $g$ remains at its plateau $n+1$. The vertical scale of $g$ is schematic.}
\label{comp:fig:ramps}
\end{figure}

\paragraph{Three functions of a data fraction.}
Let $\clp(z)=\min\{1,\max\{0,z\}\}$ and $Q=2n(n+2)$. We define
\begin{equation}
\begin{gathered}
h(t)=\clp(4t),\qquad r(t)=\clp(4t-2),\\
g(t)=(n+1)\,\clp(4t-1)+(Q-n-1)\,\clp(4t-3).
\end{gathered}
\label{comp:eq:ramps}
\end{equation}
Each of these functions varies only on certain quarters of $[0,1]$, so that a single data fraction $a_i$ serves different purposes on different quarters (Figure~\ref{comp:fig:ramps}):
\begin{itemize}
\item on $[0,\frac{1}{4}]$, the \emph{activation function} $h$ increases from zero to one; it is constant afterwards;
\item on $[\frac{1}{4},\frac{1}{2}]$, the \emph{singleton function} $g$ increases to its plateau value $n+1$;
\item on $[\frac{1}{2},\frac{3}{4}]$, the \emph{signal function} $r$ increases from zero to one, while $g$ remains constant;
\item on $[\frac{3}{4},1]$, $g$ increases to its peak $g(1)=Q$.
\end{itemize}
We call leaf $i$ \emph{active} if $a_i\ge\frac{1}{4}$, equivalently if $h(a_i)=1$. The DPGE variable of vertex $i$ is encoded as
\[
s_i=r(a_i).
\]

\paragraph{Utilities.}
For each vertex $j$, let $T_j=\{j\}\cup\{i:(j,i)\in E\}$ denote $j$ together with its out-neighbors in $G$, and define
\begin{equation}
P_j(a)=r(a_j)\prod_{i:(j,i)\in E}h(a_i),
\label{comp:eq:products}
\end{equation}
where an empty product equals one. With $D=n^2(2n+5)$, the utilities are
\begin{equation}
\begin{gathered}
U(a)=\frac{1}D\Bigl(\sum_{i=1}^n g(a_i)+\sum_{j=1}^n|T_j|\,P_j(a)\Bigr),\\
v_i(b_i)=\frac{n+2}{D}\,b_i\qquad(i\in[n]),
\end{gathered}
\label{comp:eq:utilities}
\end{equation}
where $U$ is the utility of the center and $v_i$ that of leaf $i$. Every recipient uses the genuine Shapley rule for its utility.

The center's utility consists of two kinds of terms:
\begin{itemize}
\item a \emph{singleton term} $g(a_i)$ depends on the data of leaf $i$ alone;
\item a \emph{product term} $P_j$ vanishes unless every member of $T_j$ supplies data. These terms model complementarities, and they are the only place where the edges of $G$ appear.
\end{itemize}
If all leaves in $T_j\setminus\{j\}$ are active, then $P_j(a)=r(a_j)=s_j$, so the product term carries the signal of vertex $j$.

\paragraph{Choice of constants.}
We refer to a credit at the center multiplied by $D$ as a \emph{scaled credit}. The constants play the following roles.
\begin{itemize}
\item \emph{Target $n+2$.} Each leaf attains its maximum utility $(n+2)/D$ precisely when $b_i=1$, and the corresponding scaled credit $n+2$ serves as the target for every leaf. It exceeds $n-1$, which bounds the scaled credit a leaf can receive from its in-neighbors alone; this gap will force every leaf to be active.
\item \emph{Plateau $n+1$.} The plateau of $g$ equals the target minus one, so that on the plateau the remaining unit must be supplied by the leaf's own signal and its incoming signals.
\item \emph{Peak $Q=2n(n+2)$.} This is twice the total target of all leaves, which makes full bilateral exchange with any single leaf profitable for the center.
\item \emph{Normalization $D=nQ+n^2$.} It scales all utilities into $[0,1]$ while preserving a constant Lipschitz bound.
\end{itemize}

\paragraph{Accuracy.}
For a DPGE instance with accuracy $\delta$, we request an exchange at accuracy $\eps=\delta/(4D)$.
To expose the exact mechanism, the statements below also allow $\eps=0$; their quantitative forms hold uniformly for $0\le\eps\le1/(4D)$.

\subsection{Shapley credits and validity of the instance}
\label{comp:subsec:shares}

We first verify that the intended credit formula is produced by the Shapley rule.

\begin{lemma}[Shapley shares of the instance]
\label{comp:lem:shares}
The center's Shapley share to donor $i$ is $c_{i0}(a)=\Phi_i(a)/D$, where the scaled credit is
\begin{equation}
\Phi_i(a)=g(a_i)+\sum_{j:\,i\in T_j}P_j(a).
\label{comp:eq:shares}
\end{equation}
At leaf $i$, the center receives credit $c_{0i}(x)=v_i(b_i)$ and every other donor receives zero. All utilities are continuous, monotone, lattice-supermodular, zero at the zero bundle, and bounded by one. All shares are nondecreasing in every coordinate of their recipient's bundle.
\end{lemma}
\begin{proof}
Fix the center's bundle $a$ and compute the marginals in~\eqref{eq:shapley} term by term. Since $h(0)=r(0)=0$, the product $P_j$ vanishes unless every member of $T_j$ is present. Hence, along an ordering $\pi$, the term $|T_j|P_j$ jumps from zero to $|T_j|P_j(a)$ exactly when the last member of $T_j$ arrives, and this member collects the whole term. Each member of $T_j$ is last with probability $1/|T_j|$, so each receives $P_j(a)$ in expectation, and donors outside $T_j$ receive nothing. The singleton term $g(a_i)$ changes only when $i$ arrives, so it is credited entirely to $i$. Summing over all terms and dividing by $D$ gives~\eqref{comp:eq:shares}. At a leaf, the center is the only donor that affects utility, so it receives the entire utility.

For validity, every factor of $P_j$ is a nonnegative, nondecreasing function of a distinct coordinate. A rectangular difference in two distinct coordinates is therefore the product of the two nonnegative factor increments and the remaining nonnegative factors, so $P_j$ has increasing differences. Singleton terms are lattice-modular, nonnegative sums preserve lattice supermodularity, and the leaf utilities are linear. Continuity, monotonicity, and $U(0)=0$ follow from the definitions. Since $g\le Q$, $P_j\le1$, and $|T_j|\le n$, we obtain $0\le U\le(nQ+n^2)/D=1$. Finally, every term in~\eqref{comp:eq:shares} is coordinatewise nondecreasing.
\end{proof}

A leaf $i$ belongs to $T_j$ either when $j=i$ or when $(j,i)\in E$. Consequently, if every leaf is active, then $P_j(a)=s_j$ for all $j$, and~\eqref{comp:eq:shares} becomes
\begin{equation}
\Phi_i(a)=g(a_i)+s_i+S_i.
\label{comp:eq:active-credit}
\end{equation}

\subsection{Core stability and reciprocity fix the credit targets}
\label{comp:subsec:targets}

Core stability is used only through deviations by the center and a single leaf. We say that the pair $\{0,i\}$ is \emph{$\eps$-blocking} if full bilateral exchange between the two agents improves each of them by more than $\eps$. Since the leaf utilities ignore data from other leaves, leaf $i$ earns credit only at the center. Its surplus is therefore $c_{i0}(x)-v_i$, and the center's surplus is $\sum_iv_i-U(a)$.

\begin{proposition}[Credit targets]
\label{comp:lem:targets}
At every exactly reciprocal, core-stable exchange,
\begin{equation}
b_i=1\quad\text{and}\quad\Phi_i(a)=n+2\qquad(i\in[n]).
\label{comp:eq:targets-exact}
\end{equation}
More generally, let $0\le\eps\le1/(4D)$. If the exchange is $\eps$-reciprocal and has no $\eps$-blocking center--leaf pair, then
\begin{equation}
|\Phi_i(a)-(n+2)|\le2D\eps\qquad(i\in[n]).
\label{comp:eq:targets}
\end{equation}
\end{proposition}
\begin{proof}
We prove the approximate statement; for $\eps=0$ the argument below yields $v_i=(n+2)/D$, which forces $b_i=1$, and~\eqref{comp:eq:targets} becomes~\eqref{comp:eq:targets-exact}.

Since the center receives the entire utility of each leaf as credit, its total credit equals $\sum_iv_i$. Reciprocity at the center therefore bounds its current utility:
\[
U(a)\le\sum_iv_i+\eps\le\frac{n(n+2)}{D}+\eps.
\]
Now fix a leaf $i$ and suppose that the center and leaf $i$ exchange their full data. The center then obtains utility at least $g(1)/D=Q/D=2n(n+2)/D$, an improvement of at least $n(n+2)/D-\eps$. This improvement exceeds $\eps$, because $2D\eps\le\frac{1}{2}<n(n+2)$. Leaf $i$ obtains its maximum utility $(n+2)/D$. Since the pair is not $\eps$-blocking, leaf $i$ cannot gain more than $\eps$, and hence
\[
\frac{n+2}{D}-\eps\le v_i\le\frac{n+2}{D}.
\]
Reciprocity at leaf $i$ gives $|c_{i0}(x)-v_i|\le\eps$, and therefore $-2\eps\le c_{i0}(x)-(n+2)/D\le\eps$. Multiplying by $D$ yields~\eqref{comp:eq:targets}.
\end{proof}

\subsection{The credit targets make every leaf active}
\label{comp:subsec:activation}

By~\eqref{comp:eq:shares}, the scaled credit $\Phi_i(a)$ of leaf $i$ consists of its singleton term $g(a_i)$, its own product term $P_i(a)$, and one product term $P_j(a)\le1$ for each vertex $j$ with $(j,i)\in E$. If $a_i\le\frac{1}{4}$, the first two terms vanish, and the remaining at most $n-1$ product terms cannot reach the target $n+2$. The following proposition makes this precise.

\begin{proposition}[Activation]
\label{comp:lem:activation}
Let $0\le\theta\le\frac{1}{2}$, and suppose that $|\Phi_i(a)-(n+2)|\le\theta$ for every $i$. Then $a_i>\frac{1}{4}$ and $h(a_i)=1$ for every $i$. Consequently,~\eqref{comp:eq:active-credit} holds and
\begin{equation}
|g(a_i)+s_i+S_i-(n+2)|\le\theta\qquad(i\in[n]).
\label{comp:eq:decode}
\end{equation}
In particular, every exactly reciprocal, core-stable exchange satisfies $g(a_i)+s_i+S_i=n+2$ for all $i$, and every exchange covered by Proposition~\ref{comp:lem:targets} satisfies~\eqref{comp:eq:decode} with $\theta=2D\eps$.
\end{proposition}
\begin{proof}
Suppose that $a_i\le\frac{1}{4}$ for some $i$. Then $g(a_i)=r(a_i)=0$, and in particular $P_i(a)=0$. By~\eqref{comp:eq:shares}, every remaining term of $\Phi_i(a)$ stems from an in-neighbor $j$ of $i$. There are at most $n-1$ such vertices, and $P_j\le1$ for each of them, so
\[
\Phi_i(a)\le n-1<n+2-\theta,
\]
contradicting the hypothesis. Hence $a_i>\frac{1}{4}$ and $h(a_i)=1$ for every $i$. It follows that $P_j(a)=s_j$ for every $j$, which gives~\eqref{comp:eq:active-credit}, and the hypothesis then yields~\eqref{comp:eq:decode}. The final claims follow from Proposition~\ref{comp:lem:targets}.
\end{proof}

Activation is exact even for approximate exchanges. The gap between $n-1$ and $n+2-\theta$ places every activation factor on its flat region, so no error accumulates along the products, and the only remaining discrepancy is the additive residual $\theta$ in~\eqref{comp:eq:decode}.

\subsection{Decoding the source equilibrium}
\label{comp:subsec:decode}

On the plateau of $g$, subtracting $n+1$ from~\eqref{comp:eq:decode} yields $s_i+S_i\approx1$, which is the interior case of the best response. Outside the plateau, the constant regions of $r$ implement the two boundary cases: $s_i=0$ when the incoming signals suffice, and $s_i=1$ when they vanish.

\begin{lemma}[Decoding]
\label{comp:lem:decode}
Let $0\le\eps\le1/(4D)$. For every $\eps$-reciprocal exchange with no $\eps$-blocking center--leaf pair, the vector $s$ given by $s_i=r(a_i)$ satisfies
\begin{equation}
\left|s_i-\Bigl[1-\sum_{j:(j,i)\in E}s_j\Bigr]_+\right|
\le2D\eps
\quad(i\in[n]).
\label{comp:eq:decoded-error}
\end{equation}
Consequently, every exactly reciprocal, core-stable exchange decodes to an exact DPGE equilibrium, and for $\eps=\delta/(4D)$ every such approximate exchange decodes to a DPGE solution with error at most $\delta/2$.
\end{lemma}
\begin{proof}
Let $\theta=2D\eps\le\frac{1}{2}$. By Propositions~\ref{comp:lem:targets} and~\ref{comp:lem:activation}, inequality~\eqref{comp:eq:decode} holds. Fix a vertex $i$ and distinguish three cases according to the value of $a_i$; the bounds agree at the common endpoints.

If $a_i\le\frac{1}{2}$, then $s_i=0$ and $g(a_i)\le n+1$, so the lower bound in~\eqref{comp:eq:decode} gives $S_i\ge1-\theta$. Hence $|s_i-[1-S_i]_+|=[1-S_i]_+\le\theta$. In the exact case, the in-neighbors supply at least one unit and the best response is zero. Note that the leaf is nevertheless active and thus contributes to the product terms of other vertices.

If $\frac{1}{2}\le a_i\le\frac{3}{4}$, then $g(a_i)=n+1$, and~\eqref{comp:eq:decode} reduces to $|s_i+S_i-1|\le\theta$. Since $s_i\ge0$ and the map $t\mapsto[t]_+$ is nonexpansive,
\[
|s_i-[1-S_i]_+|=|[s_i]_+-[1-S_i]_+|\le|s_i-(1-S_i)|\le\theta.
\]
In the exact case, the leaf's own signal fills precisely the deficit left by its in-neighbors.

If $a_i\ge\frac{3}{4}$, then $s_i=1$ and $g(a_i)\ge n+1$, so the upper bound in~\eqref{comp:eq:decode} gives $0\le S_i\le\theta$. Hence $|s_i-[1-S_i]_+|=\min\{1,S_i\}\le\theta$. In the exact case, the incoming signal vanishes and the best response is one.
Substituting $\eps=0$ or $\eps=\delta/(4D)$ yields the two consequences.
\end{proof}

In short, the credits lie within $\theta=2D\eps$ of their targets, activation is exact for $\theta\le\frac{1}{2}$, and each decoded signal lies within $\theta$ of its best response. Since $D$ is polynomial in $n$, the accuracy $\eps=\delta/(4D)$ is inverse polynomial.

\subsection{Hardness}
\label{comp:subsec:hardness}

It remains to verify that the constructed instance and its Shapley oracle can be computed in polynomial time, and that the large coefficient $Q$ does not create a large Lipschitz constant after normalization.

\begin{theorem}\label{comp:thm:hardness}
For some inverse-polynomial accuracy $\eps=1/\operatorname{poly}(N)$, finding an $\eps$-reciprocal and $\eps$-core-stable exchange is $\PPAD$-hard for $N$ agents with continuous, monotone, lattice-supermodular utilities and genuine Shapley shares. Hardness holds even when:
\begin{enumerate}
\item only center--leaf exchanges affect utility, the center is the only agent with a nonlinear utility, and every leaf utility is linear;
\item utilities take values in $[0,1]$ and vanish at the zero bundle;
\item all utilities and shares are $6$-Lipschitz in $\ell_\infty$ and exactly computable at rational inputs in polynomial time;
\item core stability is required only against two-agent coalitions.
\end{enumerate}
\end{theorem}
\begin{proof}
Given a DPGE instance on $n$ vertices at its hard inverse-polynomial accuracy $\delta$, construct the instance~\eqref{comp:eq:utilities} with $N=n+1$ agents and set $\eps=\delta/(4D)$, which is inverse polynomial in $N$ since $D=n^2(2n+5)$. Lemma~\ref{comp:lem:shares} establishes the required utility properties and the Shapley formulas. An $\eps$-core-stable exchange has no $\eps$-blocking center--leaf pair, so Lemma~\ref{comp:lem:decode} converts every $\eps$-reciprocal, $\eps$-core-stable exchange into a DPGE solution with error at most $\delta/2\le\delta$. Since the argument uses only center--leaf deviations, the same conclusion holds under property~(4).

\emph{Construction and evaluation.} The constants $Q$ and $D$ have $O(\log n)$ bits, each of $g,h,r$ is a piecewise-linear function with a constant-size description, and~\eqref{comp:eq:utilities} has $O(n+|E|)$ terms and factor occurrences. Formula~\eqref{comp:eq:shares} provides an exact share oracle on the entire domain, not only on bundles that encode equilibria: one computes each product $P_j(a)$ once, adds it to the scaled credit of every member of $T_j$, adds the singleton terms, and divides by $D$. This requires $O(n+|E|)$ arithmetic operations on rationals of polynomial bit length, since each product has at most $n$ factors. In particular, no estimation of Shapley values is needed.

\emph{A constant Lipschitz bound.} The functions $h$ and $r$ are $4$-Lipschitz, and $g$ is $4Q$-Lipschitz. A product of $k$ factors with values in $[0,1]$, each $4$-Lipschitz in a distinct coordinate, is $4k$-Lipschitz in $\ell_\infty$. Using $|T_j|\le n$ and $Q=2n(n+2)$, we obtain
\begin{equation}
\begin{aligned}
\Lip_\infty(U)
&\le\frac{4nQ+4\sum_j|T_j|^2}{D}
\le\frac{8n^2(n+2)+4n^3}{n^2(2n+5)}\\
&=\frac{12n+16}{2n+5}<6.
\end{aligned}
\label{comp:eq:lipschitz}
\end{equation}
The same bound holds for every center share. For bundles $a,b$, let $\ell=a\wedge b$ and $z=a\vee b$. Coordinatewise monotonicity of the shares and Shapley efficiency give
\[
\begin{aligned}
|c_{i0}(a)-c_{i0}(b)|&\le c_{i0}(z)-c_{i0}(\ell)\\
&\le U(z)-U(\ell)\le6\norminf{a-b}.
\end{aligned}
\]
The leaf utilities and shares are $(n+2)/D$-Lipschitz, and $(n+2)/D<1$.
\end{proof}

\begin{corollary}\label{comp:cor:complete}
For monotone lattice-supermodular utilities with Shapley shares, polynomial-time utility and share evaluation, and a constant Lipschitz bound, the exchange problem at inverse-polynomial accuracy is $\PPAD$-complete. In particular, it admits no FPTAS unless $\PPAD\subseteq\FP$.
\end{corollary}
\begin{proof}
Hardness follows from Theorem~\ref{comp:thm:hardness}. Membership follows from the general $\PPAD$ upper bound of Akrami et al.~\cite{Akrami2025}, which applies when utilities and shares can be evaluated efficiently and $L/\eps$ is polynomially bounded. An FPTAS would solve the hard instances at inverse-polynomial accuracy in polynomial time.
\end{proof}

\FloatBarrier
\section*{AI Methodology Statement}
\phantomsection
\addcontentsline{toc}{section}{AI Methodology Statement}
Both the algorithm and the hardness result were developed through active interactions with ChatGPT 6.0 (Astra). The authors verified the mathematical arguments, simplified the proofs, and prepared the final exposition, including a technical overview explaining the key ideas behind the advances over prior work. ChatGPT 6.0 also assisted with writing and editing the manuscript. The authors take full responsibility for the correctness and presentation of the results.

\phantomsection
\addcontentsline{toc}{section}{References}
\newcommand{\etalchar}[1]{$^{#1}$}

\end{document}